# Generation of Fine Particles with Specified Characteristics

A.N. Ishmatov, V.V. Elesin[2], A.A. Trubnikov [1], S.P. Ogorodnikov [2]

1: Institute for Problems of Chemical & Energetic Technologies
of the Siberian Branch of the Russian Academy of Sciences (IPCET SB RAS), Russia
2: Join Stock Company Federal Research & Production Center ALTAI, Russia

**Abstract**
The research is devoted to the generation of fine particles (particles size below 5 μm) and gas-droplet flows with specified characteristics for a wide range of scientific problems. The aspects of aerodynamic fine atomization and the effects of atomizing gas density, gas velocity and mass flow rate, and liquid film thickness on the droplet formation are investigated. The hypothesis of highly efficient utilization of secondary droplets from a coarse polydisperse flow to produce fine particles is suggested and experimentally confirmed. A prototype device to implement the idea of separating the desired droplets' fraction from a primary polydisperse flow was developed. In the case of fine liquid atomization, the developed spraying system enabled an increase in the gas-droplet flow concentration. The possibility of producing particles with different dispersiveness and morphology by employing the methods of spray separation and dilute solution atomization is demonstrated. The criteria of the system settings to generate the droplet flow with specified characteristics were determined. A Laval nozzle design is suggested for liquid atomization. A series of numerical experiments showed that a significant decrease in gas pressure (density) at the nozzle outlet provides an increase in the gas flow rate up to 734 m/s. The numerical estimation showed that the Laval nozzle proposed can improve the fine atomization efficiency by ~ 36%.

**Introduction**
Fine particles are widely used in the modern world. The special attention is paid to the generation of particles and gas-droplet flows with a high surface activity (high surface area of particles). The high surface area of particles ensures highly effective neutralization of toxic aerosols, gases, and smoke, and reduces the negative environmental impact when agricultural chemicals are used.
Ultrafine particles have found their widespread application in medicine for pulmonary drug delivery as a noninvasive administration of drugs. Features of the pulmonary drug delivery application are characterized by the depth of penetration and deposition of ultrafine drug particles in the lungs. The introduction of new drug delivery technologies requires the development of effective methods and tools to generate preparations with desired characteristics, as well as the study of delivery principles and mechanisms. The elaboration of new approaches for the production of substances having particles with specified morphology as well as the investigation of properties and kinetics of such particles is promising.
There is a major problem in the industry associated with improving the manufacture efficiency of ultrafine materials to design new types of energetic constituents for fuels and of electric power sources.
Current trends in the ecologization and reduction of pollutions caused by the sources of smoke and gas emissions make it important to study the interaction between ultrafine particles and the environment.
The problem we deal with is the generation of fine particles with specified characteristics (dispersion, concentration, flow rate) for a broad range of scientific problems.

**Research problem formulation**
To produce submicron nanosized particles, the method of dispersing dilute solutions containing volatile solvents (spray drying) has widely been used. The effectiveness of this approach is due to the efficiency of liquid atomization. For the generation of ultrafine droplets, the most widely applied method is aerodynamic atomization [1, 2].
The fine and ultrafine liquid atomization is due to the complex influencing factors such as strong turbulence, air resistance, cavitation, etc. [1, 3]. The liquid disintegration occurs in the near-nozzle region chaotically and highly irregularly [1, 4, 5]. At the first stage, there is formed a conical liquid film on the nozzle edge. The aerodynamic gas flow exposure causes formation of unstable surface waves, which leads to the liquid film sheet breakup and spray formation [1, 4, 6]. Schematically, the liquid disintegration can be represented as shown in fig. 1. Dombrowski and Johns [7], on the basis of analytical relations, the analysis of aerodynamic instability, and disintegration of viscous liquid films, suggested the following expression for estimating the diameter of ligaments formed upon the breakup of a planar liquid film:

* Corresponding author: ishmatoff@rambler.ru

$$d_L = 2\left(\frac{4}{3f}\right)\left(\frac{K_N^2\sigma^2}{\rho_g\rho_L U_f^4}\right)^{1/6} \cdot \left(1 + 2.6\mu_L\left(\frac{K_N\rho_g^4 U_f^8}{72\rho_L^2\sigma^5}\right)^{1/3}\right)^{1/5}, \quad (1)$$

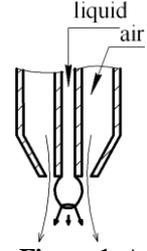

**Figure 1.** A schematic showing liquid film sheet breakup mode in aerodynamic atomization

where $d_L$ is the diameter of ligaments; $f$ is the constant (for large Weber Numbers, We>>1, $f$=12 [8]; $K_N$ is the "nozzle parameter"; $\sigma$ is the liquid surface tension; $\rho_g$ and $\rho_L$ are the gas and liquid densities, respectively; $U_f$ is the velocity of gas flowing around the liquid film; $\mu_L$ is the liquid viscosity. The "nozzle parameter" is determined by the design of the nozzle and, in a general case, can be expressed as [8]: $K_N \approx t_s h$, where $t_s$ is the time required for disintegration of the liquid film; $h$ is the film thickness. For an incompressible flow [9, 10], the following relation is suggested to estimate parameter $K$:

$$K_N = h^2 \cos^3\theta / U_L, \quad (2)$$

where $\theta$ is the cone semi-angle of the liquid flow; $U_L$ is the liquid velocity at the atomizer tip. For an aerodynamic atomizer, $U_L = U_f$. Thus, for large Weber Numbers and substituting relations (1) and (2), we will have:

$$d_L = 0.9614\cos\theta\left(\frac{h^4\sigma^2}{\rho_g\rho_L U_f^4}\right)^{1/6} \cdot \left(1 + 2.6\mu_L\cos\theta\left(\frac{h^2\rho_g^4 U_f^7}{72\rho_L^2\sigma^5}\right)^{1/3}\right)^{1/5}. \quad (3)$$

According to expression (3), the dispersiveness can be altered by changing the liquid film thickness ($h$), the gas velocity ($U_f$), and the gas density ($\rho_g$). The estimation of the influence of these parameters on the liquid atomization is given below.

**Mathematical model**

For the experimental study, we chose a standard aerodynamic nozzle shown in fig. 2.
To estimate the flow characteristic, we built a mathematical model of the intra-chamber channel with a variable cross-section. For building the model, we used non-stationary differential equations of gas dynamics in a one-dimensional case, expressing the law of mass conservation (4), momentum (5), energy (6), and equation of state (7) written for the internal energy.

$$\partial \rho_g S / \partial t + \partial \rho_g S U / \partial x = 0, \quad (4)$$

$$\partial \rho_g S U / \partial t + \partial s\,(\rho_g U^2 + P)/\partial x = 0, \quad (5)$$

$$\partial \rho_g S\,(e + U^2/2)/\partial t + \partial\left[\rho_g U\,(e + P/\rho_g + U^2/2)\right]/\partial x = 0, \quad (6)$$

$$e = P/[\rho_g(k-1)], \quad (7)$$

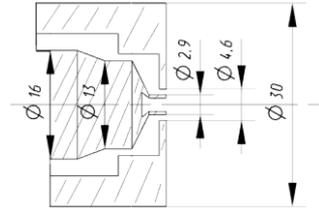

**Figure 2.** A nozzle drawing

where $U$, $P$, $e$ are the velocity, the pressure, and the energy of the gas flow; $S$ is the cross-section area of the channel; $t$ is the time; $x$ is the coordinate; $k$ is the adiabatic exponent. The inlet boundary conditions for a given inlet gas pressure, $P_{in}$, are calculated by the relations:

$$\rho_{in} = P_{in}/(R_v T), \qquad G = S \cdot P \cdot \sqrt{R_v T/k^{-1}} \cdot ((k+1)/2)^{-\frac{k+1}{k-1}}, \qquad U_{in} = G/(\rho_{in} \cdot S_{in}),$$

where $R_v$ is the gas constant; $T$ is the gas temperature; $G$ is the gas mass flow rate through the nozzle; $S_{in}$ is the inlet cross-sectional area.
The outlet boundary conditions are calculated for known drag parameters on one of the preceding nodes (because drag parameters characterize the gas state in this node for any process) [11].

$$P_{out} = P_{I1} \cdot ((k+1)/2)^{-\frac{k}{k-1}}, \qquad \rho_{out} = \rho_{I1} \cdot ((k+1)/2)^{-\frac{1}{k-1}}, \qquad U_{out} = G/(\rho_{I1} S_{out}),$$

where $I1$ is the penultimate node.
To implement the calculation of equations (4-6), a grid was built and a finite-difference scheme (8-10) suggested by S. Godunov [12] was employed that is robust and stable when solving similar problems.

$$(\rho_g \cdot S)_i^{j+1} = (\rho_g \cdot S)_i^j - \tau/l \cdot (M_i^j - M_{i-1}^j), \quad (8)$$

$$(\rho_g \cdot U \cdot S)_i^{j+1} = (\rho_g \cdot U \cdot S)_i^j - \tau/l \cdot (I_i^j - I_{i-1}^j) - \tau P_i^j/l \cdot (S_i^j - S_{i-1}^j), \quad (9)$$

$$\left[S \cdot \rho_g \cdot \left(\frac{P}{(k-1)\cdot\rho_g}\right) + \frac{U^2}{2}\right]_i^{j+1} = \left[S \cdot \rho_g \cdot \left(\frac{P}{(k-1)\cdot\rho_g}\right) + \frac{U^2}{2}\right]_i^j - \frac{\tau}{l}\cdot(E_i^j - E_{i-1}^j), \quad (10)$$

where $i$ is the node number along the axis of the nozzle; $j$ is the number of the time iteration; $l$ is the step size (length divided by the number of nodes); $\tau$ is the time step, which is calculated according to the Courant condition ($\tau = K \cdot l / c$, where $K$ is the Courant number, $c$ is the sound speed in the gas); $S$ is the cross-sectional area in the node $i$; $M$, $J$, $E$ are the flows of mass, momentum, and energy calculated by the formulae:

$$M = S \cdot \rho'_g \cdot P', \quad J = S \cdot \left[ P' + \rho'_g \cdot (U')^2 \right], \quad E = S \cdot \left[ \rho'_g \cdot U' \cdot P' / (k-1) \cdot \rho'_g + U'^2 / 2 \right],$$

where $\rho'_g$, $P'$, $U'$ are the density, pressure, and velocity at the contact discontinuity between two adjacent grid nodes, which are calculated according to the method proposed by S. Godunov [12]. Using the above methods, the dependence of the gas density at the outlet orifice on the inlet gas pressure was numerically investigated (fig. 3).

It was also found that, for the chosen nozzle and steady flow, the inlet gas pressure had almost no effect on the gas flow velocity and the volume flow rate. The outlet flow velocity is limited to 350 m/s; the volume flow rate is $2.46 \cdot 10^{-3}$ m$^3$/s.

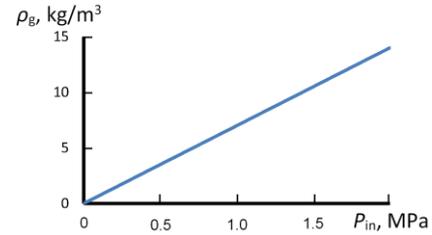

**Figure 3.** Influence of the inlet gas pressure on the gas density at the outlet orifice

This phenomenon can be explained as follows: with increasing gas pressure, the density and the mass flow increase at the same time by a proportional value, thus giving $G_V = G / \rho_{out} = const$.

**Influence of the gas jet density on liquid atomization**

In this work, the following parameters were used: distilled water for spraying, liquid flow rate ~ 0.003 kg/s, operating gas pressures from 0.2 MPa to 2.0 MPa, which corresponded to the change in the outlet gas density from 1.4 kg/m$^3$ to 14 kg/m$^3$ (calculation results).

For the numerical experiment, the thickness of the liquid film was estimated using (11) [10] and reverse numerical calculations of (3) for a known Sauter mean diameter (SMD).

$$h = 0.5 \left[ d_0 - \left( d_0^2 - 4 m_L / \pi \rho_L U_L \right)^{\frac{1}{2}} \right], \quad (11)$$

where $d_0$ is the final discharge orifice diameter; $m_L$ is the mass flow rate of the liquid.

For measurements of the spray characteristics, a Malvern Spraytec laser analyzer (Malvern Instruments, Malvern, UK) [13] was employed. The calculation results and experimental data are presented in fig. 4.

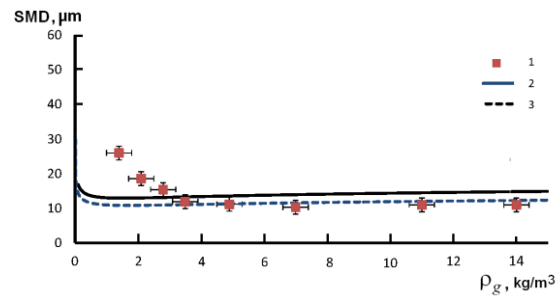

**Figure 4.** Influence of the gas density on the mean droplet size: 1 – experiment; 2 – calculation, $h$=190 μm (Eq.(11)); 2 – calculation, $h$=150 μm (Eq. (3))

The results of numerical and experimental studies showed that the influence of the gas density (from 3 kg/m$^3$ to 14 kg/m$^3$) on the dispersiveness was insignificant. The experiment showed that the mean droplet size decreased with increasing gas density up to 3 kg/m$^3$. This phenomenon can be attributed to a non-stationary nozzle injection mode. Further increase in the gas density leads to no variation in the dispersiveness, which is in agreement with the calculations.

**Influence of the gas velocity on liquid atomization**

A numerical estimation was performed to study the possibility of improving the efficiency of liquid atomization. The thickness of the liquid film is considered to be 150 μm and gas density – 4 kg/m$^3$, which corresponds to the given nozzle and the experimental conditions. The calculation results are given in fig. 5.

The above analysis of the atomizer showed that the outlet gas velocity did not depend on the inlet gas pressure and was limited to the value of ~350 m/s. The numerical estimation of the gas velocity effect on the mean droplet size demonstrated that an increase in the gas velocity (above 350 m/s) led to a decrease in the mean droplet size. To optimize the atomization (increasing gas velocity), we propose a simple design of the Laval nozzle (fig. 6).

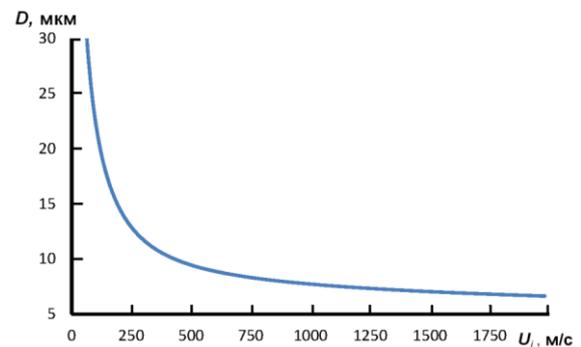

**Figure 5.** The gas velocity versus the mean droplet size ($h$=150 μm)

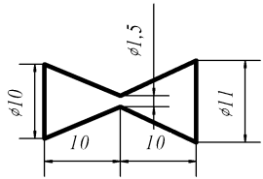

**Figure 6.** A schematic of the Laval nozzle

The Laval nozzle has a high efficiency of the conversion of compressed gas energy to the flow velocity. Considering the following parameters, i.e. volume gas flow rate from $2.5 \cdot 10^{-3}$ m³/s to $5.0 \cdot 10^{-3}$ m³/s and the inlet gas pressure of 0.5 MPa, we performed a series of numerical experiments. The numerical calculation revealed that the simple nozzle of Laval type gave an increment in the outlet gas velocity from 354 m/s to 734 m/s owing to a significant pressure decrease in the nozzle throat (~0.01 MPa). The numerical estimation showed that the Laval nozzle used would allow an improvement in the ultrafine atomization performance by ~ 36% (the average SMD decreased from 11.26 μm to 8.24 μm).

**Influence of the film thickness on liquid atomization**

Theoretical and experimental results of the estimation of the film thickness influence on atomization are presented in fig. 7. The theoretical estimation was performed by numerical computation of expression (3). The experimental estimation was performed for the following conditions: $m_L$ ~0,003 kg/s, inlet gas pressure ($P_{in}$) ~0.5-0.7 MPa, $h$=150 μm. In the experiments, the film thickness was varied by changing the liquid mass flow rate. The comparison between the film thickness and the mass flow rate of the liquid was made in accordance with relation (11) for the assumption that the liquid velocity at the atomizer tip is independent of the liquid mass flow rate and equal to the outlet gas velocity (~ 350 m/s). Thus, fig. 7 shows that the decrease in the film thickness leads to a decrement in the average droplet size in the spray.

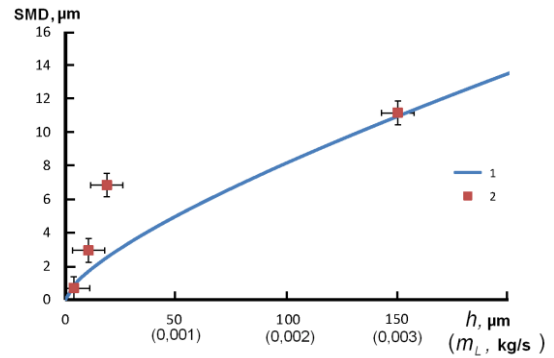

**Figure 7.** The film thickness as a function of the mean droplet size: 1 – calculation; 2 – experiment

Obviously, there is a restriction on the film thickness value, caused by surface tension, viscosity, as well as surface wettability of the atomizer tip. In the experiments, the reduction of the flow rate allowed us to increase the amount of the fine droplet fraction in the gas-droplet flow (results are also shown in fig. 8):

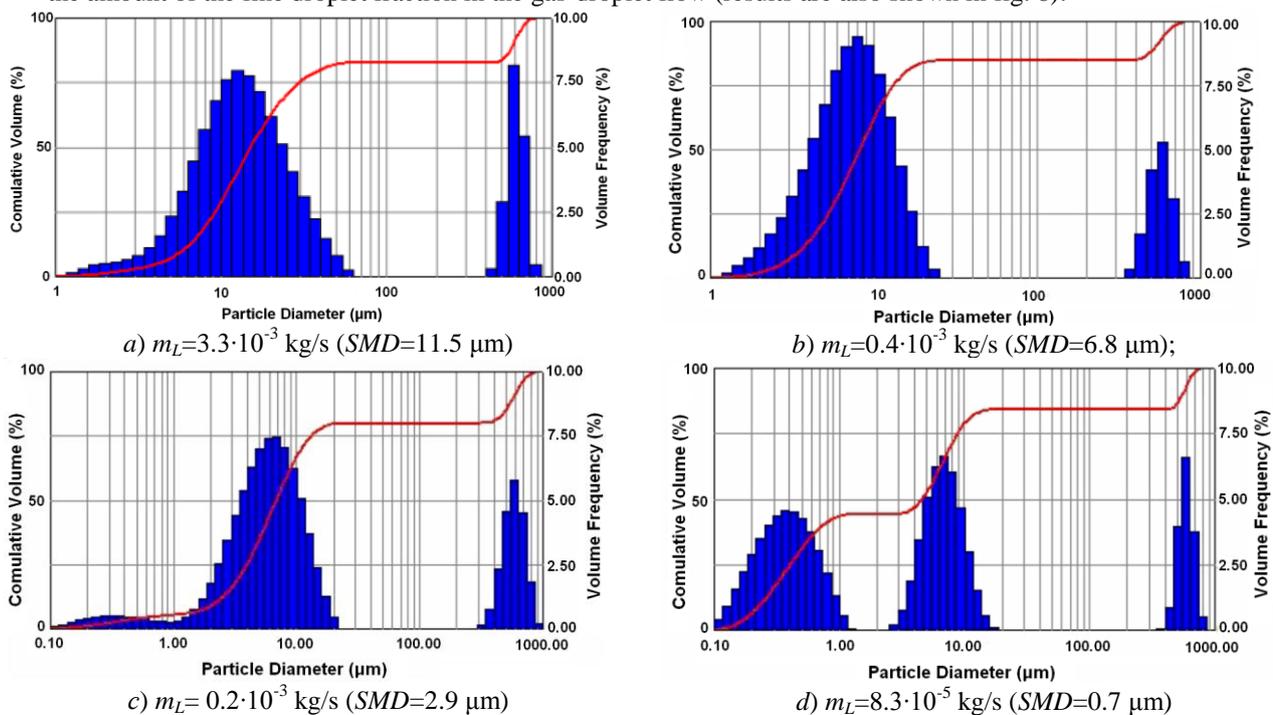

*a)* $m_L$=3.3·10⁻³ kg/s (*SMD*=11.5 μm)                                     *b)* $m_L$=0.4·10⁻³ kg/s (*SMD*=6.8 μm);

*c)* $m_L$= 0.2·10⁻³ kg/s (*SMD*=2.9 μm)                                     *d)* $m_L$=8.3·10⁻⁵ kg/s (*SMD*=0.7 μm)

**Figure 8.** The droplet mass distribution function for different liquid flow rates at the atomizer tip

Fig.8d illustrates the possibility of generating submicron droplets using the atomizer with a minimized liquid flow rate at the atomizer tip. At this point, to atomize 100 g of liquid into droplets with SMD=6.8 μm (fig.8b), the gas volume of ~ 1.3 m³ is required; for SMD=2.9 μm, the gas volume of ~2.5 m³ is necessary. The droplet concentration in the gas flow was ~0.077 kg/m³ (SMD=6.8 μm) and ~0.038 kg/m³ (SMD=2.9 μm), respectively (here, we mean the droplet concentration in the gas flow at atmospheric pressure).

It was noted that, for the maximum liquid flow rate ($m_L$=3.3·10$^{-3}$ kg/s), the percentage by volume of droplets below 10 μm in the main flow was ~30%; the percentage by volume of droplets below 6 μm was ~12%.

Thus, to produce 100 g of droplets less than 10 μm in size, it is required to atomize 303 g of liquid into droplets with SMD=11.5 μm; the required gas volume is 0.5 m$^3$ (the mass of all droplets is 303 g and 100 g (30 %) is the mass of droplets less than 10 μm in size); to produce 100 g of droplets less than 6 μm, it is needed to atomize 833 g of liquid into droplets with SMD=11.5μm and the required gas volume is ~0.9 m$^3$ (the mass of all droplets is 833 g and 100 g (12 %) is the mass of droplets less than 6 μm).

Thus, it can be assumed that the utilization of secondary fine and ultrafine droplets from the coarse polydisperse flow to generate fine droplets is more effective than primary fine liquid atomization.

**Spray separation**

The method of spray separation is widely used to produce fine droplets [14-17]. In this paper, to implement the idea of separating the desired droplet fraction, a primary polydisperse flow is passed into a separator as a spirally rolled-up pipe (fig. 9). The separation occurs in the field of centrifugal forces when droplets move into the pipe.

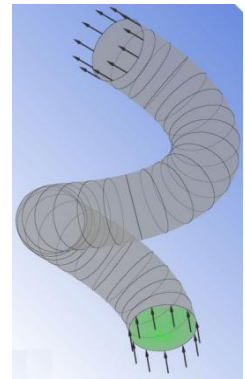

The separation is adjusted by changing the parameters: flow rate, number of turns, and pipe radius. Experimentally, by using a flexible corrugated pipe, we determined the parameters of the separator. Further, on the basis of the obtained data, we designed a separator system from polyvinyl chloride.

The experimental study revealed the restriction on the use of the separator in the performance mode (operating gas pressures above 0.4 MPa)—this is due to the increase in the flow rate in the pipe and the decrease in the separation quality. Results of the study devoted to solving this problem will be published in the nearest future.

In terms of maximum efficiency, the following parameters were chosen: operating gas pressure – 0.4 MPa, liquid flow rate ~ 0.003 kg/s, pipe diameter – 0.1 m, diameter of the rolling-up is about 0.3 m, the number of turns – 1; height – 0.5 m. To stabilize the evaporation of the droplets in the flow, we used a 10% aqueous solution of glycerol. The designed system allows generating the droplet flow with SMD=2.7 μm, droplet concentration – 0.055 kg/m$^3$, and production capacity ~0.167·10$^{-3}$ kg/s (the data are shown in fig. 10).

**Figure 9.** A schematic of the separator

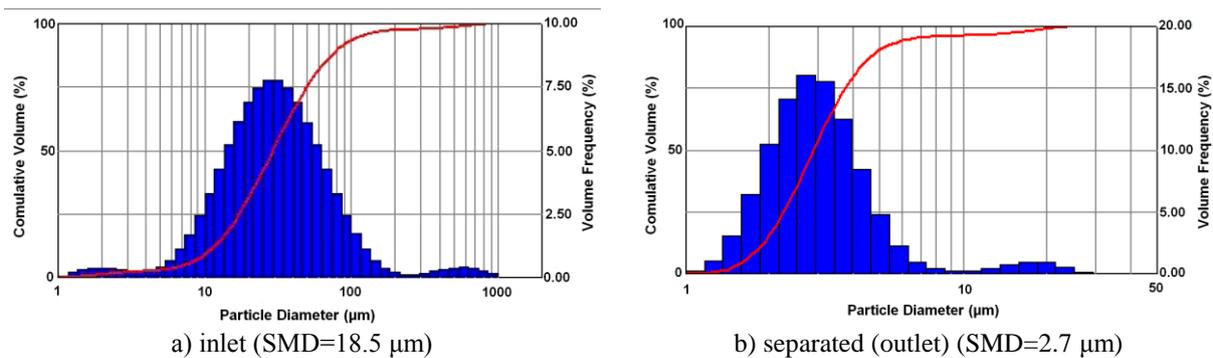

a) inlet (SMD=18.5 μm)    b) separated (outlet) (SMD=2.7 μm)
**Figure 10.** The droplet mass distribution function

The experimental results showed that the aerodynamic atomization method combined with the method of spray separation, in contrast to the "direct" setting of the aerodynamic atomizer, allows producing an ultrafine gas-droplet flow with a single-mode droplet mass distribution function close to the gamma function. The designed system enabled an increase in the gas-droplet flow concentration in the case of fine liquid atomization (SMD~2.7 μm) from 0.038 kg/m$^3$—for a single aerodynamic atomizer – to 0.055 kg/m$^3$—for the separator mode. The obtained experimental results confirm the hypothesis of highly efficient utilization of secondary droplets from the polydisperse flow to generate fine droplets.

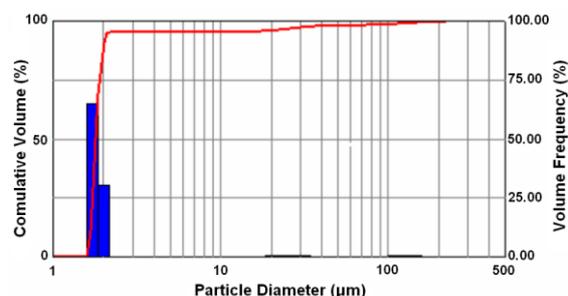

**Figure 11.** The monodisperse droplet flow generation

The possibility of producing a fraction of fine droplets with certain dispersiveness is shown in fig. 11b.

Obviously, to obtain submicron droplets at the outlet of the separator, they must be contained in the droplet flow at the inlet. The content of desired fine droplets in the inlet flow results in their concentration at the outlet. To produce ultrafine particles, one may use the method of dispersing dilute solutions containing volatile solvents (spray drying). By changing the concentration of the solutions, particles with specified sizes can be produced [2, 18].

Also, by controlling the evaporation and crystallization processes, particles with different morphologies can be achieved [19]. As an example, one can consider the morphology of the particles formed as a result of the droplets evolution in a high-speed, impulse dispersed flow (fig. 12) [20-22].

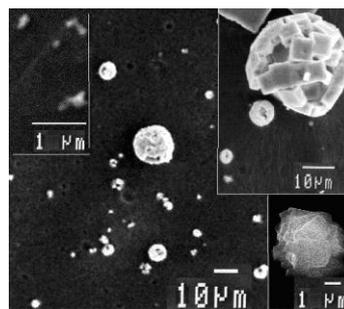

**Figure 12.** The morphology of the particles formed in a high-speed, impulse dispersed flow [21]

The non-stationary conditions and high evaporation rate of droplets lead to the formation of particles with different morphologies [19].

Thus, the possibility of generating particles with different dispersiveness and morphology by using the spray separation method and the method of dispersing dilute solutions have been demonstrated.

**Conclusions**

As a consequence, the generation of fine particles and gas-droplet flows with specified characteristics for a wide range of scientific problems has been investigated. The aspects of aerodynamic fine atomization and the effects of atomizing gas density, gas velocity and mass flow rate, and liquid film thickness on the droplet formation have been studied.

As a result of gas-dynamic calculations, we have found that the standard atomizer is low efficient because of the limitation of the outlet gas velocity (less than 350 m/s). For further investigation, we proposed a design in the form of a Laval nozzle. A series of numerical experiments have shown that a significant decrease in the gas pressure (density) at the nozzle outlet provides an increase in the gas flow rate up to 734 m/s. The numerical estimation has shown that the Laval nozzle utilization will allow an improvement in the ultrafine atomization efficiency by ~ 36 %.

The hypothesis of high efficient utilization of secondary droplets from the polydisperse flow to generate fine particles have been suggested and experimentally confirmed. A prototype device implementing the idea of separating the desired droplet fraction from the primary polydisperse flow has been developed. The experimental results showed that the aerodynamic atomization method, coupled with the method of spray separation, in contrast to the "direct" setting of the aerodynamic atomizer, makes it possible to produce an fine gas-droplet flow with a single-mode droplet mass distribution function close to the gamma function.

In the case of fine liquid atomization, the designed spray system enabled the increase in the gas-droplet flow concentration. The possibility of efficiently generating the fine fraction of droplets with certain dispersiveness has been demonstrated. The criteria of the system settings for the generation of a droplet flow with specified characteristics have been determined.

The research results have many technological and scientific applications, e.g., the urgent task of improving the productivity and efficiency of technologies for the synthesis of ultrafine materials.

**Acknowledgements**
The research was partially supported by the RFBR, research project No. 12-08-31282 mol_a.

**Nomenclature**
$c$ is the speed of sound in the gas;
$d_L$ is the diameter of ligaments;
$d_0$ is the final discharge orifice diameter;
$E$ is the flows of energy;
$e$ is the energy of the gas flow;
$G$ is the gas mass flow rate;
$G_V$ is the gas volume flow rate;
$h$ is the film thickness;
$I1$ is the penultimate node
$J$ is the flows of momentum;
$K$ is the Courant number;
$K_N$ – is the "nozzle parameter";
$k$ is the adiabatic exponent.
$l$ is the step size (length divided by the number of nodes);
$M$ is the flows of mass;

$m_L$ is the mass flow rate of liquid;.
$P$ is the gas pressure;
$P_{in}$, $P_{out}$ are the inlet and outlet gas pressure;
$R_v$ is the gas constant;
$S$ is the cross section area; $S_{in}$ is the inlet sectional area.
$T$ is the gas temperature;
$t_s$ is the time required for disintegrated of the liquid film;
$U$ is the gas velocity;
$U_f$ is velocity of gas flowing the liquid film;
$U_L$ is the liquid velocity at the atomizer tip;
$x$ is the coordinate;

$\theta$ is the cone semi-angle of liquid flow

$\mu_L$ is the liquid viscosity.

$\rho_g$ and $\rho_L$ are the gas densities and liquid, respectively;

$\sigma$ is the liquid surface tension;

$\tau$ is the time step, which is calculated according to the Courant condition ( $\tau = K \cdot l / c$ );